# Tuning Magnetic and Optical Properties in $Mn_xZn_{1-x}PS_3$ Single Crystals by the Alloying Composition

Adi Harchol[1#], Shahar Zuri[1#], Esther Ritov[1], Faris Horani[1,2], Miłosz Rybak[3], Tomasz Woźniak[3], Anna Eyal[4], Yaron Amouyal[5], Magdalena Birowska[3], Efrat Lifshitz[1*]

1. Schulich Faculty of Chemistry, Solid State Institute, Russell Berrie Nanotechnology Institute, Helen Diller Quantum Information Center, Technion, Haifa 3200003, Israel
2. Department of Chemistry, University of Washington, Seattle, Washington 98195-1700, United States
3. Institute of Theoretical Physics, Faculty of Physics, University of Warsaw, 02-093 Warsaw, Pasteura 5, Poland
4. Department of Physics, Technion, Haifa 3200003, Israel
5. Department of Material Science and Engineering, Technion, Haifa 3200003, Israel

*The corresponding author: ssefrat@technion.ac.il

# Equal contribution

**Abstract**

The exploration of two-dimensional (2D) antiferromagnetic (AFM) materials has shown great promise and interest in tuning the magnetic and electronic properties as well as studying magneto-optical effects. The current work investigates the control of magneto-optical interactions in alloyed $Mn_xZn_{1-x}PS_3$ lamellar semiconductor single crystals, with the Mn/Zn ratio regulating the coupling strength. Magnetic susceptibility results show a retention of AFM order followed by a decrease in Néel temperatures down to ~ 40% Mn concentration, below which a paramagnetic behavior is observed. Absorption measurements reveal an increase in bandgap energy with higher Zn(II) concentration, and the presence of Mn(II) d-d transition below the absorption edge. DFT+U approach qualitatively explained the origin and the position of the experimentally observed mid band-gap states in pure $MnPS_3$, and corresponding peaks visible in the alloyed systems $Mn_xZn_{1-x}PS_3$. Accordingly, emission at 1.3 eV in all alloyed compounds results from recombination from a $^4T_{1g}$ Mn(II) excited state to a hybrid p-d state at the valence band. Most significant, temperature-dependent photoluminescence (PL) intensity trends demonstrate strong magneto-optical coupling in compositions with x > 0.65. This study underscores the potential of tailored alloy compositions as a means to control magnetic and optical properties in 2D materials, paving the way for advances in spin-based technologies.

**Introduction**

The renewed interest in two-dimensional (2D) layered materials inspired by the discovery of graphene,[1] has opened a new paradigm in science and technology.[2] These materials consist of nearly atomistic layers held together by weak van der Waals (vdW) interactions, enabling the isolation of individual atomic layers.[3] Recently, 2D magnetic materials have garnered considerable interest, stemming from their promising potential in modern memory devices and spintronics applications.[4–6] Within these materials, the unpaired spins of adjacent magnetic ions are mutually coupled to form a ferromagnetic (FM) or antiferromagnetic (AFM) arrangement below the Curie ($T_C$) or the Néel ($T_N$) temperature, respectively. The existence of intrinsic FM in 2D vdW compounds has been unveiled in several materials, such as $CrI_3$, $Cr_2Ge_2Te_6$, and $Fe_3GeTe_2$, where FM order has been detected even within a single-layer.[7–10]

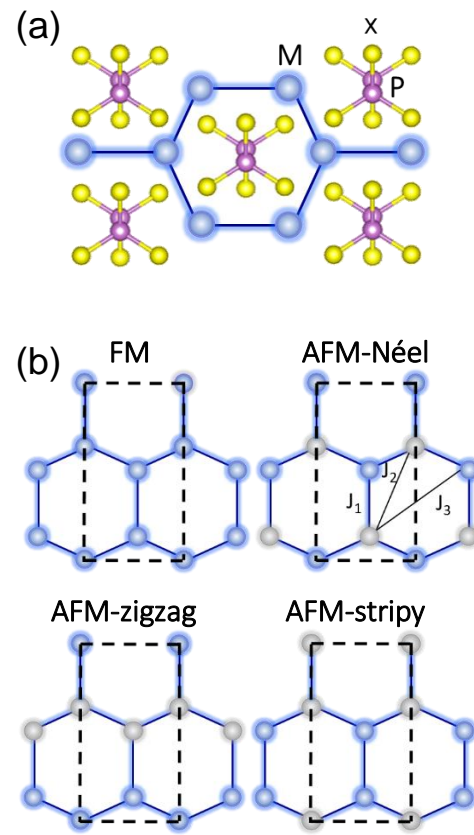

**Figure 1.** (a) Illustration of an $MPX_3$ layer. The blue, purple and yellow balls represent the metal, phosphorus and chalcogen ions, respectively. (b) Magnetic configurations: FM, AFM-Néel, AFM-zigzag, and AFM-stripy. Spin up and down are represented by blue and grey balls, respectively. Rectangular unit cell is indicated by black dashed lines. The antiferromagnetic Néel configuration highlights the exchange coupling of the first three nearest neighbors.

Conversely, the interest in 2D AFM materials is relatively new, and despite being in its infancy, their intriguing merits, such as low stray magnetic fields and a THz spin-flip frequency,[11] have already make them ideal materials for spin-transistors, spin-filters, and photodetectors.[11–16] A classical family of 2D AFM compounds are the transition metal phosphorus tri-chalcogenides, represented by the chemical formula $MPX_3$ (M= first-row transition metals, P=phosphorus, X=S, Se). In these compounds, a single layer consists of a network of shared-edge $[MX_6]^{2+}$ and $[P_2X_6]^{4-}$ octahedral units (in a ratio of 2:1). Metal ions across the layer are arranged in a honeycomb array with $P_2$ located at the hexagon centers (**Figure 1(a)**). This metal framework facilitates AFM coupling between metal ions, aligning their spins either perpendicular (Ising or anisotropic Heisenberg) or parallel (XY) to the lamellar planes.[17,18] The long-range magnetism in $MPX_3$ compounds primarily arises from exchange interactions between the unpaired spins of nearest neighbors (NNs) metal ions. Such exchange interactions enable the formation of AFM-Néel (e.g.,

MnPS$_3$),[19] AFM-zigzag (e.g., FePS$_3$, NiPS$_3$)[18,19] and AFM-stripy magnetic configurations, as illustrated in **Figure 1(b)**.

The spin-exchange interactions in MPX$_3$ materials are described by a simple Heisenberg Hamiltonian (**Eq. 1**), typically involving interactions up to the third NN:[6,20–22]

(Eq.1)
$$H = \frac{1}{2}\sum_{i,j} J_1 \vec{S}_i \cdot \vec{S}_j + \frac{1}{2}\sum_{i,j} J_2 \vec{S}_i \cdot \vec{S}_j + \frac{1}{2}\sum_{i,j} J_3 \vec{S}_i \cdot \vec{S}_j$$

In this equation, *i* and *j* represent the metal sites, while $J_1$, $J_2$, and $J_3$ are the exchange coupling constants between the first, second, and third NNs, respectively, as denoted in the AFM-Néel configuration in **Figure 1(b)**). $S_{i,j}$ denotes the spin operator at site *i,j*. Note that for systems involving crystalline anisotropy, dipolar interactions, or Zeeman coupling, the Hamiltonian should incorporate additional terms to fully describe the magnetic behavior.[6,20,23]

The MPX$_3$ vdW compounds also show semiconductor properties, which bring an additional merit related to the correlation between the magnetic and optical properties, as evidenced by several reports.[24–31] For instance, Raman spectroscopy has uncovered spin-phonon coupling below the Néel temperature in FePS$_3$ and MnPS$_3$,[31–33] and magnon-phonon hybridization in MnPSe$_3$.[34] Exciton-magnon coupling has been observed in bulk MnPS$_3$,[35] and a linear optical emission in FePS$_3$ and NiPS$_3$, related to the AFM-zigzag arrangement[26] or directionality,[36] have also been reported. In addition, recent DFT-based calculations unveiled that in MPX$_3$ systems, the polarization type light coupled to optical transitions is dependent on the AFM spin arrangement.[37,38] Pump-probe spectroscopy has detected a slowdown in spin dynamics near the Néel temperature of FePS$_3$.[39] More recently, a PL study on MnPS$_3$ has identified a near-infrared (NIR) transition correlated with the Néel ordering.[30]

Despite these initial observations regarding AFM MPX$_3$ compounds, a fundamental understanding of the correlation between long-range spin-exchange interactions and optical properties has remained elusive. To address this question, this work implemented a dilution of the magnetic ions within the metal honeycomb lattice with diamagnetic constituents, aiming to modulate the strength of the first NN exchange coupling and to further examine whether the AFM configuration is preserved via distant interactions. The materials under investigation include Mn$_x$Zn$_{1-x}$PS$_3$ alloyed compounds, where a MnPS$_3$ matrix host accommodates variable amounts of diamagnetic Zn(II) ions. MnPS$_3$ is an ideal model system due to its negligible spin-orbit coupling, and being a direct bandgap semiconductor.[40] In addition, Mn(II) ions possess a ground state spin of 5/2 and an AFM-Néel arrangement below 78 K,[41] with the spins oriented nearly perpendicular to the honeycomb plane (with slight tilt of ~8°).[42] The choice of Mn and Zn, giving their similar ionic radii, minimizes internal strains and aggregations. As an added value, this study uncover the potential of variable alloy composition as a control knob for the magnetic configuration (more details in the results section).

This study explores the intricate interplay between optics and magnetism in bulk 2D $Mn_xZn_{1-x}PS_3$ alloyed compounds ($0.37<x\leq1$) by comparing magnetic susceptibility with absorption, variable temperature steady-state PL, and transient-PL measurements. The optical measurements uncover the role of Mn(II) d-orbitals in the optical transitions, which serve as footmarks for drastic changes upon the change in composition. These optical observations, in agreement with the magnetic susceptibility data, reveal a reduction in $T_N$ as x decreases, thus supporting a correlation between the long-range magnetic order and optical properties. Furthermore, the study shows a retention of the AFM characteristic down to a composition of x=0.4. Finally, experimental findings were corroborated by DFT+U calculations, while taking a brute force approach to consider various magnetic phases in $Mn_xZn_{1-x}PS_3$ systems. The developed magnetic model offers insights into the magnetic ground state of the alloys, exchange parameters up to third NN, and the Néel temperature for each composition, in agreement with experimental results.

**Results and Discussion**

**Synthesis and characterization of $Mn_xZn_{1-x}PS_3$ crystals.** Bulk single crystals of $Mn_xZn_{1-x}PS_3$ were synthesized using a vapor-transport method, starting with stoichiometric amounts of the respective elements, as detailed in Ref.[43]. The crystallographic structure of the pristine sample was verified by X-ray diffraction (XRD), revealing a monoclinic crystal structure with C2/m space group. A representative XRD pattern of a $MnPS_3$ single crystal is shown in the supplement information (SI), **Figure S1(a)**. The concentration and distribution of the metal elements in various $Mn_xZn_{1-x}PS_3$ single crystals were verified by a high-resolution scanning electron microscopy (HR-SEM) combined with an energy dispersive X-ray (EDX) analysis. Based on these measurements, a set of samples were selected for the current study comprising the Mn(II) fractions (x) of 0.37, 0.50, 0.65, 0.77, and 0.88. For convenience, they are labeled henceforth as Mnx.

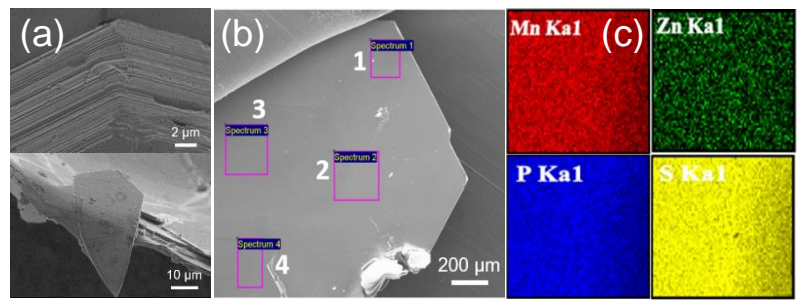

**Figure 2.** (a) HR-SEM images of $MnPS_3$ crystals (top and bottom). (b) SEM image of the Mn0.77 crystal. The pink rectangles represent the areas where elemental analysis was performed. (c) SEM elemental mapping of zone '4' in (b) displaying a uniform atomic distribution.

**Figures 2(a)** and **(b)** exhibit HR-SEM images of the $MnPS_3$ and Mn0.77 crystals, respectively, confirming their high crystallinity, stacking arrangement, and well-defined surface morphology. The pink squares in **Figure 2(b)**, mark sub-areas for which explicit elemental analysis were recorded by EDX. Furthermore, **Figure 2(c)** presents EDX maps depicting the distribution of Mn, Zn, P, and S elements in area '4' of panel **(b)**, each represented by a different color. These maps confirm, within the resolution limits of the EDX experiments (~ 20 nm), a uniform distribution of these elements. A representative EDX spectrum of Mn0.77 at area '1' is provided in the SI, **Figure S1(b)**, while the concentration of metal elements in all marked areas in panel **(b)** are summarized in **Table S1**. These results underline the crystals' high purity and consistent stochiometric proportions across different sub-areas. The impact of composition was also exposed by Raman spectroscopy. **Figure S1(c)** shows a set of Raman spectra, presenting gradual changes in the spectral range below 200 cm$^{-1}$, evolving from Mn-rich to Zn-rich content. This spectral window is dominated by the symmetric and antisymmetric modes of the metal vibrations within the lamellar plane (labeled as P2 and P3 in **Figure S1(c)**). The gradual change in the metal-vibration modes further indicates a uniform distribution of metal elements in the alloyed compounds. An extensive Raman study on this samples was recently published in Ref.[41]. A homogeneous distribution of the elements was further supported by theoretical calculations of the bond distances at low temperatures within a unit cell (SI, **Table S2**). In MPS$_3$ materials, the layer skeleton is largely dictated by the bipyramid [P$_2$S$_6$]$^{4-}$ unit size, where the metal ions with smaller ionic radii compared to the P-P bond ($r_{Mn(II)}$ = 0.83 Å, $r_{Zn(II)}$ = 0.74 Å, P-P = 2.22 Å for MnPS$_3$), exhibit partial freedom within the [MS$_6$]$^{2+}$ unit. **Table S2** reveals a minor reduction (less than 1%) in the unit cell size with increasing Zn content, which corresponds to a slight decrease in the bipyramid [P$_2$S$_6$]$^{4-}$ unit size. These theoretical evaluations suggest a reduced potential for aggregation, strain or vacancies in the as-grown bulk single crystals, although not entirely eliminating them.

**Magnetic properties of bulk Mn$_x$Zn$_{1-x}$PS$_3$.** The impact of the diamagnetic ion on the magnetic properties of Mn$_x$Zn$_{1-x}$PS$_3$ was exploited by examining the magnetic susceptibility ($\chi$) versus temperature, while investigating single crystals with different compositions. The experiments involved the use of a superconducting quantum interference device (SQUID) operating at temperatures ranging between 4K to 300K and under a weak external magnetic field (1 kOe) oriented parallel (∥) or perpendicular (⊥) to the a-b plane of the crystals. Note that MnPS$_3$ possesses an AFM-Néel configuration with an easy magnetic vector oriented along the tilted c-crystallographic axis, viz., would have proportionate projections along the two monitored directions.

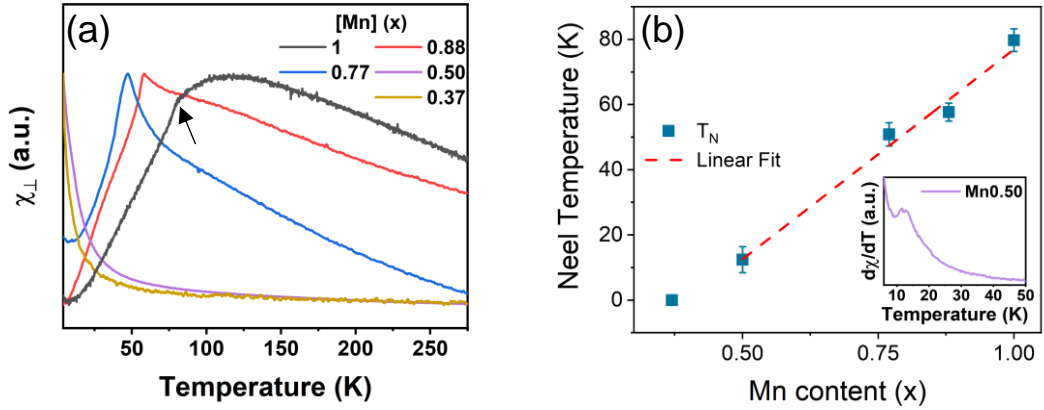

**Figure 3.** (a) perpendicular magnetic susceptibility of the alloyed crystals under a magnetic field of 1 kOe. (b) The Néel temperature of the alloyed crystals, derived from the derivative of the parallel magnetic susceptibility with temperature. Inset: the derivative of the susceptibility of the Mn0.50 sample.

**Figure 3(a)** shows temperature-dependent susceptibility plots under a perpendicular magnetic field ($\chi_\perp$) for samples indicated by the legend. The complementary data recorded under parallel magnetic field ($\chi_\parallel$) are displayed at the SI, **Figure S2(a)**. For the pristine $MnPS_3$, a decrease in temperature elevates the $\chi_\perp$ intensity up to ~120K, after which it stabilizes until reaching the Néel temperature at $T_N$ = 79.2±2.4 K (black arrow in **Figure 3(a)**). The regime between 300K to 120K corresponds to a paramagnetic phase, while the interval from 120K to $T_N$ is characterized as a frustrated magnetic stage,[44] which persists until the spins are locked into an ordered AFM-Néel configuration at $T_N$. Although the AFM nature remains at temperatures below $T_N$, a decline in the $\chi_\perp$ intensity is noticeable, presumably due to the canting of spins toward the a-b plane.[45,46]

The $\chi_\perp(T)$ trends of the Mn0.88 and Mn0.77 samples share similarities to pristine $MnPS_3$, including a decrease in susceptibility below $T_N$. Nevertheless, a noticeable shift in their $T_N$ to lower temperatures is evident with a dependence on the Mn fractions (justified in the simulations below) accompanied by a blurred frustration region.

**Figure 3(a)** provides insights into the magnetic behavior of the Mn0.50 and Mn0.37 samples. The $\chi_\perp(T)$ trend of the Mn0.50 crystal veils a fine structure detail, but a plot of $d\chi/dT$ (inset in **Figure 3(b)**) displays a discernible hump associated with an AFM-Néel point, around ~12 K, along with a contribution of para-magnetism. The $\chi_\perp(T)$ trend of the Mn0.37 sample exhibits a total loss of AFM character and is compatible with a paramagnetic behavior (following the Curie-Weiss law). Overall, the $T_N$ values extracted from the SQUID measurements and their dependence on the Mn content are plotted in **Figure 3(b)**. This plot reveals a linear relationship, in agreement with the anticipated trend after gradually incorporating Zn(II) into the host crystals.[47,48] Ultimately, the AFM magnetic order persists for Mn(II) fractions higher than x=0.4±0.1, as determined by extrapolating the linear fit. This finding aligns well with previous work, where the critical composition was found to be x=0.46±0.03.[48]

**Optical properties of bulk $Mn_xZn_{1-x}PS_3$.** The room temperature absorption spectra of $MnPS_3$ and Mn0.37 samples are shown in **Figure 4(a)**. The spectra comprise fundamental absorption edges at 2.79 eV (for $MnPS_3$) and 2.95 eV (for Mn0.37), along with several well-defined sub-bandgap peaks, denoted as $E_i$. Noting that pristine $ZnPS_3$ possesses a bandgap of 3.5 eV,[49] thus explaining the band-edge shift toward higher energy in Mn0.37. The $E_i$ peaks were fitted by Gaussian functions (the fit and the used parameters are provided in **Figure S3** and **Table S3**, respectively), with maximal energies consistent with those reported in the literature.[50] These $E_i$ peaks are attributed to the spin and parity forbidden d-d transitions of a localized Mn(II) ion at a octahedral site, from the ground state, $^6A_{1g}$, to three different excited states: $^4T_{1g}$, $^4T_{2g}$, and ($^4E_g$, $^4A_{1g}$), where the latter contains two degenerate levels.[50] The energies of the 3d atomic-like Mn(II) excited states are in good agreement with those previously reported for Mn-doped semiconductors.[51] In the present case, the third transition, $E_3$ ($^6A_{1g}\rightarrow(^4E_g, ^4A_{1g})$), is hidden by the room temperature absorption edge of $MnPS_3$, yet it is visible in the Mn0.37 crystal spectrum. Upon incorporating Zn(II) into the lattice, both $E_1$ and $E_2$ bands in the Mn0.37 spectrum are shifted towards lower energies by 40 meV and 80 meV, respectively. Notably, the intensities of the $E_i$ transitions in **Figure 4(a)** are relatively strong with respect to the band-edge scale, reflecting a partially allowed transition as a result of p-d hybridization in the valence band (VB).[40] The involvement of d-orbitals in the absorption spectra and the p-d hybridization in the VB are further verified by DFT+U calculations, as will be discussed below.

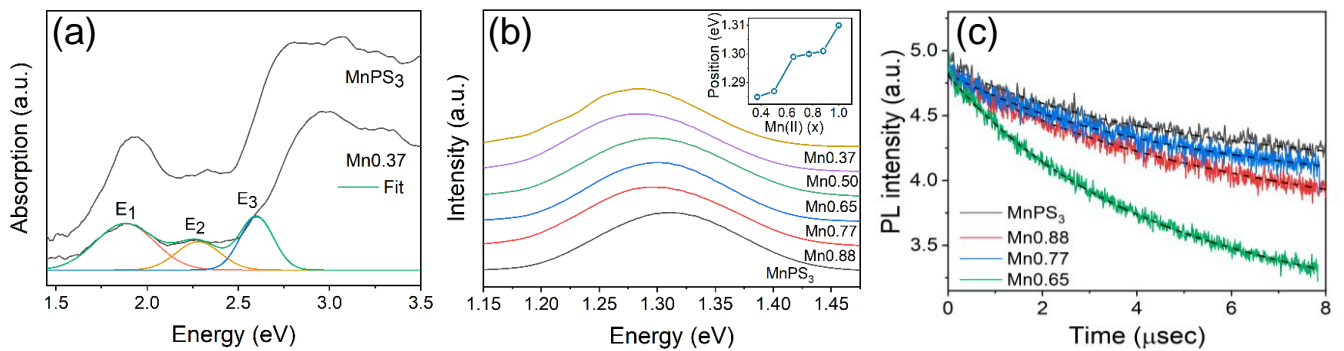

**Figure 4.** (a) Absorption spectra with spectral deconvolution of $MnPS_3$ and Mn0.37 recorded at room temperature. (b) PL spectra of $Mn_xZn_{1-x}PS_3$ obtained at a temperature of 5 K, while exciting with a CW 405 nm laser. Inset: the position of the PL peak as a function of the Mn(II) content, derived from a Gaussian fit. (c) PL decay curves of $Mn_xZn_{1-x}PS_3$ recorded at a temperature of 5 K while exciting with a pulsed laser of 450 nm wavelength. The best fit is depicted by the dashed black line.

**Figure 4(b)** depicts a set of PL spectra of bulk $Mn_xZn_{1-x}PS_3$ single crystals acquired at 5K. The samples were excited using a 405 nm continuous-wave (CW) laser. The PL spectra comprise a single broad band with a full-width-at-half-maximum (FWHM) of about 140 meV, which gradually shifts towards the near-infrared regime with the increase of Zn(II) content (see inset). The remarkable red shift of 30 meV is close to the energy shift observed in the absorption $E_1$ peak upon diamagnetic doping, confirming that the emission originates from the transition between the Mn(II) $^4T_{1g}$ excited state and the hybrid p-d state in the VB (from now on referred to as $^4T_{1g} \rightarrow {}^6A_{1g}$(VB)). The total PL Stokes shift of ~600 meV in the emission spectra with respect to $E_1$ is attributed to the low dielectric screening in thin semiconductors.[37,52] It is important to pay attention that excitation with sub-bandgap energy showed a significantly weaker emission band (SI, **Figure S4**), implying that the PL emission is sensitized via an absorption above the band-edge, followed by a single or a cascade of relaxation processes into the Mn(II) excited states.

**Figure 4(c)** presents a set of PL-decay curves for the samples listed in the legend. These curves were best fitted by bi-exponential functions (black dashed lines), revealing two different decay times for each sample, ranging from 96 to 169 μsec for the longer component and 3 to 3.5 μsec for the shorter component (SI, **Table S4**). The long component dominates the decay in all samples. However, its contribution decreases with the decrease of the Mn(II) content. Essentially, PL decay times at the μsec range support the occurrence of a $^4T_{1g} \rightarrow {}^6A_{1g}$(VB) transition characterized by a partially allowed nature. This contrasts with the typical msec decay in an atomistic Mn(II) $^4T_{1g} \rightarrow {}^6A_{1g}$ transition.[51] The discrepancy is related to the fact that the Mn(II) $^6A_{1g}$ ground state is involved in p-d hybridization at the edge of the VB.[40] The faster component in the decay curve is more likely associated with a non-radiative recombination process.[53]

**Figure 5(a,c)** depicts the electronic structures of pure MnPS$_3$ and ZnPS$_3$ bulk crystals, respectively. These were calculated using the DFT+U (PBE+U) and meta-GGA ($E_{XC}$=mBJ) exchange-correlation functionals as described in the Methods. Note that U=1.8 eV was applied to Mn 3d states based on previously published ARPES measurements,[54] ensuring the correct position of the 3d states within the valence bands relative to the valence band maximum (VBM). It is assumed that the absorption onset in MnPS$_3$ originates from valence bands deeper than the top of the VB. Namely, the fundamental absorption edge (neglecting excitons) occurs between the top of the green region (consisting of p states) in **Figure 5(a)** to the bottom of the CB (marked in red), resulting in a fundamental electronic band gap of $E_{gap}^{MnPS3}$ =2.7 eV. While the electronic band gaps calculated by these approaches still underestimate the actual values, the calculated difference between ZnPS$_3$ and MnPS$_3$ band gaps ($E_{gap}^{ZnPS3}$-$E_{gap}^{MnPS3}$ = 3.2 eV - 2.7 eV = 0.5 eV) (see **Figures 5(a,c)**) closely matches the experimental band gaps differences observed in these bulk materials measured at low temperatures (3.5 eV - 3 eV = 0.5 eV).[49] The choice of the onset of the absorption edge is further strengthened by the selection rules favoring transitions between different types of orbitals (Δl= ±1). Additionally, the agreement between the theoretical $E_i$ values and the experimental $E_i$ peaks, which are located below the absorption edge

(**Figure 4(a)**), confirms the chosen onset of the absorption edge. Hence, the $E_i$ peaks are attributed to all direct transitions occurring between the bands VB1 → CB1 ($E_1$) and VB2 → CB1 ($E_2$) at each of k-points, resulting in an energy window equals to 0.5 eV (the blue region) visible in the projected density of states (PDOS) (**Figure 5(a)**, right). The $E_3$ peak can be attributed to the transition from VB1 to the conduction bands marked in red in region IV (yellow region). Importantly, the former energy window (0.5 eV) reflects the energy difference between the peaks $E_2$ and $E_1$ and aligns well with the energy difference of the experimental peaks $E_2 - E_1 = 2.5 - 1.9 = 0.6$ eV. Moreover, the energy itself matches. That is, assuming that the $E_1$ transition occurs from VB1 to CB1 at K high symmetry point (see **Figure 5(a)**), its energy equals 1.9 eV, consistent with the maximum peak position of the experimental value ($E_1 = 1.9$ eV). Based on the band structure of $MnPS_3$, $ZnPS_3$ and the crystal field theory of metal ion with electronic configuration $3d^5$ in octahedral geometry with ground state ($^6A_{1g}$) and excited states ($^4T_{1g}$, $^4T_{2g}$, $^4E_g$) (for more details see Figure 3.9 in Ref.[55]), a schematic diagram illustrating an oversimplified picture of atomic 3d $Mn^{2+}$ states embedded in the p states of P and S is constructed in **Figure 5(b)**. This diagram indicates possible optical transitions related to d-d transitions. Particularly, within the crystal field approach in octahedral coordination, the $E_1$, $E_2$, and $E_3$ transitions correspond to the electronic transitions $^6A_{1g} \rightarrow ^4T_{1g}$, $^6A_{1g} \rightarrow ^4T_{2g}$, $^6A_{1g} \rightarrow ^4E_g$, respectively. Note that the relative positions of the $e_g$ and $t_{2g}$ states are schematically depicted in **Figure 5(b)** based on the main contributions of the 3d states visible in the PDOS of $MnPS_3$. Future investigations may employ alternative methods such as Maximally Localized Wannier Functions to provide crystal field and exchange splitting parameters. **Figure 5(d)** displays the PDOS of the studied alloys. The PDOS plots uncover a dominance of Mn d-orbitals at the CB, and hybridization between Mn(d) and S(p) orbitals (blue region) at the VBM. More specifically, the Mn 3d states generate tails close to the band-edges, acting as band-gap traps, in contrast to the electronic PDOS of pure $ZnPS_3$. Note that the $E_i$ peaks attributed to the Mn(II) d-d transitions are visible in all alloys with any concentration of Mn atoms within the fundamental electronic band gap, except for pure $ZnPS_3$. It is important to note the mixed nature of the Mn-atomistic ground states at the VBM by the p-d hybridization, which supposedly bestows a partial allowance for the otherwise forbidden d-d transitions. This reason also shortens the radiative lifetime with respect to that of a pure d-d transition. Furthermore, the presence of vacancies, if existing in a minor extent, would have been blurred by the d-orbital tails,[56] contradicting the observation of a mono-Gaussian component at the emission band (SI, **Figure S5(a)**).

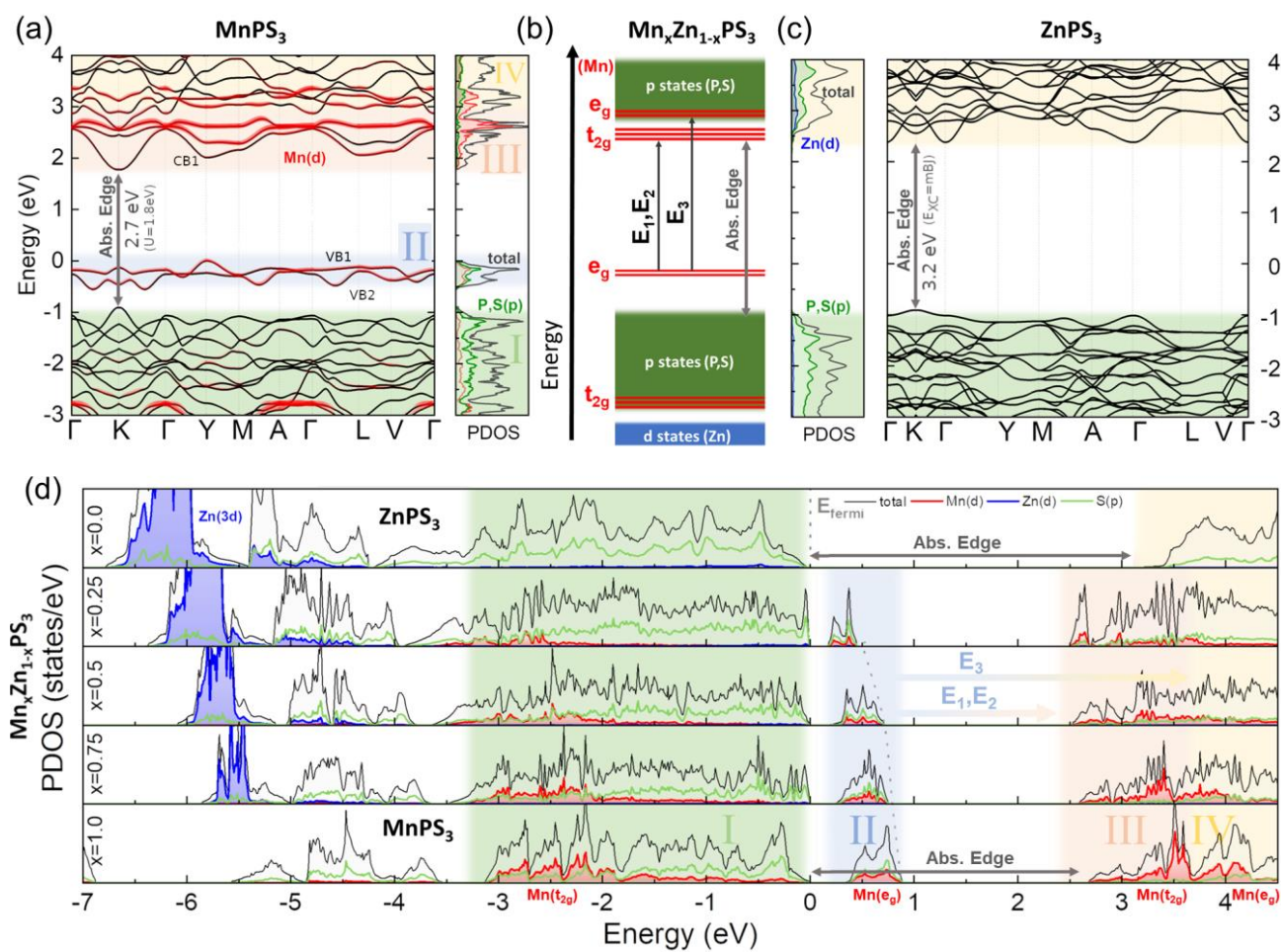

**Figure 5.** Electronic structure for the (a) MnPS$_3$ and (c) ZnPS$_3$ bulk materials. (b) Schematic diagram of optical transitions in Mn$_x$Zn$_{1-x}$PS$_3$ crystals, showing an oversimplified picture of atomic 3d states of Mn$^{2+}$ in an octahedral crystal field with t$_{2g}$ states (triple horizontal lines) and e$_g$ states (double horizontal lines) partially embedded in the p states of S and P atoms. (d) Projected density of states (PDOS) of Mn$_x$Zn$_{1-x}$PS$_3$ crystals with different Mn-content (x). Note that all the PDOS are aligned in respect to the onset of absorption edge (neglecting the excitonic effects). The p states of P are present in the entire region of the plotted CB and below the top of region I in the VB, and due to its small contribution, they are not separately plotted. All bands are occupied up to E$_{fermi}$ level marked by the grey dotted line. The abbreviation "Abs. Edge" denotes the absorption edge. In the case of the PDOS of ZnPS$_3$, the PBE approach with a rigid shift to match the band gap obtained in the mBJ approach was used.

**Temperature-dependent PL of bulk Mn$_x$Zn$_{1-x}$PS$_3$.** The validation of the optical recombination mechanisms within Mn$_x$Zn$_{1-x}$PS$_3$ lays the foundation for exploring their temperature dependence and correlation with magnetism. **Figure 6(a)** displays a color map of the temperature-dependent PL spectrum of a pristine MnPS$_3$ single crystal, exposing an intensity increase up to 50 K with a slight blue shift in energy, followed by a decline in the intensity close to the Néel temperature (dashed line in **Figure 6(a)**). A corresponding integrated PL intensity versus temperature plot is shown in the inset, extracted from fitting the PL bands by a mono-Gaussian function (SI, **Figure S5(a)**). The increase in intensity up to 50 K, as depicted in **Figure 6(a)**, stems from thermal excitation of the $^4T_{1g}$ state,[56,57] followed by a complete quenching of the emission when approaching 120 K. Note that the blue shift in the PL energy (**Figure S5(b)**) with increasing temperature was observed previously in $Mn^{2+}$-doped nanocrystals,[57,58] and attributed to the decrease in crystal field and the thermal activation of phonons. Additionally, slightly above 50 K, a knee point in the inset plot designates the start of the intensity drop. This point, labeled hereon as $T_K$ (the knee temperature), was determined through the derivative of the curve (SI, **Figure S5(c)**) to be 72.4±6.2 K for $MnPS_3$. The quenching of the intensity above $T_K$ is ascribed to the activation of a non-radiative process facilitated by phonons, thus following an Arrhenius relation as shown by the red line in the inset (more information is provided in section 8 of the SI). The temperature-dependent integrated PL intensity profiles of the alloyed crystals are depicted in **Figure 6(b)**, where the $T_K$ points were extracted in a similar manner as for the $MnPS_3$ and are marked by the black arrows. **Figure 6(c)** represents the dependence of the $T_K$ values versus the fraction of Mn (x) by the green circles. These are compared with the $T_N$ points extracted from the magnetic susceptibility (blue squares), along with a linear fit (red dashed line). Interestingly, the green symbols closely coincide with $T_N$ down to x ~ 0.65. However, they strongly deviate from the red line for x < 0.65. The observation in **Figure 6(c)** shows a strong correlation with the SQUID measurements, when the latter exposed a weak AFM behavior for Mn0.50, albeit a loss of AFM order for Mn0.37 (**Figure 3**). It is worth noting that temperature-dependent PL measurements were performed at two distinct locations on the crystals to assess the repeatability of the results. The data from both measurements exhibit consistent trends in intensity variation with temperature and yield identical knee temperatures (**Figure S6**).

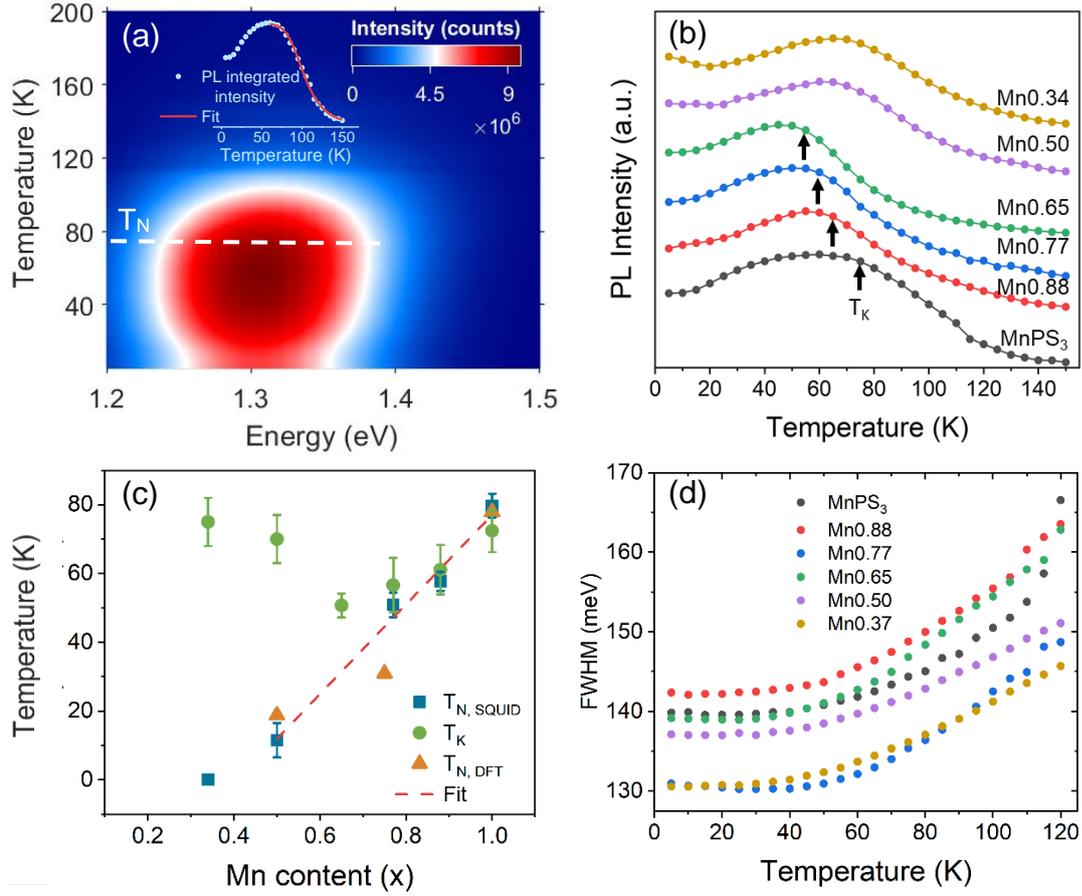

**Figure 6.** (a) A color map displaying the temperature-dependent PL of MnPS$_3$ under 405 nm laser excitation. inset: the temperature dependent of the PL integrated intensity of MnPS$_3$. (b) Temperature-dependent of the integrated PL intensity of the alloyed crystals. (c) Comparison of the Néel temperatures of the alloyed compounds derived from SQUID measurements (blue squares), the knee temperature obtained from the TD-PL measurements (green circles), and the Néel temperatures calculated from DFT simulations (orange triangles). A linear fit for the Néel temperature (extracted from SQUID) is provided (dashed red line) to guide the eye. (d) Temperature-dependent FWHM of the PL emission of the alloyed crystals.

The FWHM of the PL emission's dependence on temperature is another way to follow the magnetic influence on optical events. **Figure 6(d)** depicts a set of FMHM(T) plots for various compositions of Mn$_x$Zn$_{1-x}$PS$_3$ crystals. The trend of these curves is dictated by electron-phonon interactions, described by the relation given in **Eq. 2**:[59]

(Eq. 2) $$FWHM(T) = \Gamma_0 + \Gamma_{Ac} \cdot T + \frac{\Gamma_{LO}}{exp\left(\frac{E_{LO}}{k_B T}\right) - 1}$$

Here, $\Gamma_0$ is the broadening at 0 K, $\Gamma_{Ac}$ represents the broadening produced by acoustic phonons and $\Gamma_{LO}$ denotes the broadening produced by optical phonons with energy $E_{LO}$. The values used in the simulations were based on a recent Raman spectroscopy study on similar Mn$_x$Zn$_{1-x}$PS$_3$ crystals.[43] For a starting point, to minimize the number of variables in the simulations, The LO phonon energy, $E_{LO}$, was matched to the energy of the Raman mode P3 (155 cm$^{-1}$ for MnPS$_3$), due to its involvement with the Mn$^{2+}$ motion[60,61] and to a magnetic ordering.[31,43,62] The fit and variables are presented in the SI, **Figure S8,** and **Table S6**, exposing the following trends: (i) the values of $E_{LO}$ show a gradual change from the Mn-rich compositions to those enriched with Zn. Specifically, for Mn0.88 and Mn0.77 there is a substantial agreement between the used LO frequencies and the P3 vibrational mode of MnPS$_3$. Nonetheless, for Mn0.65 down to Mn0.37, the best fit values of $E_{LO}$ were smaller, gradually shifting towards lower energies, eventually matching the P3 mode of ZnPS$_3$ (130 cm$^{-1}$). (ii) The LO phonon coupling term, $\Gamma_{LO}$, decreases gradually, from the Mn-rich (115 meV) to the Zn-rich (50 meV), and (iii) the coupling of the photogenerated carriers and acoustic phonons, $\Gamma_{Ac}$, is negligible. The high correlation observed between the P3 vibrational mode and the photogenerated carriers in samples with x > 0.65 is in agreement with the integrated PL intensity results, demonstrating a coupling to the Néel temperature for x > 0.65. Therefore, it becomes evident that the optical events are strongly influenced by the spin-phonon coupling, the Mn content, and its magnetic arrangement. while Mn0.50 shows AFM behavior, its weak magneto-optical coupling is evident, probably due to the negligible exchange interactions as well as spin-phonon coupling.

The distinct behavior of the Mn0.50 sample was also distinguished while following the polarization of light in Mn$_x$Zn$_{1-x}$PS$_3$ crystals. Representative polar plots of a few different crystals (recorded at 5K) are shown in **Figure S9**. Unlike other samples that showed a lack of polarization, the Mn0.50 sample exhibited elliptical polarization, which decreases near its Néel temperature of ~ 12 K. Specifically, the polarization along the main ellipsoid axis coincides with a crystallographic direction. DFT+U calculations (discussed in length in the following section) verify a magnetic phase transition in the Mn0.50 sample from AFM-Néel to AFM-stripy. Moreover, these calculations reveal the presence of nearly four degenerate AFM-stripy phases within a unit cell. This can lead to a selective linear polarization along a preferred orientation, resembling to the linear polarization previously observed in NiPS$_3$ and FePS$_3$, albeit along a zigzag direction.[26,36] Nevertheless, the stochastic nature of Zn-Mn alloying in the Mn0.50 sample likely results in the coexistence of stripy and Néel configurations, leading to the observed elliptical pattern. The mechanism guiding oscillating dipoles by a magnetic directionality is an intriguing topic that would be further elaborated elsewhere.

**DFT calculations of Mn$_x$Zn$_{1-x}$PS$_3$.** The magnetic state of the alloyed compounds was elucidated using DFT+U calculations. An average magnetic phase model was implemented to consider various magnetic configurations (AFM-Néel, AFM-zigzag, AFM-stripy, FM) within the Mn$_x$Zn$_{1-x}$PS$_3$ alloyed compounds. While this analysis is applicable across the entire range of x, the focus was primarily on specific values (x = 1.00, 0.75, and 0.50) for efficiency. A comprehensive description of the magnetic phases and corresponding calculation details can be found in the methods section.

The analysis assumed a random distribution of Zn cations within the honeycomb lattice. DFT+U calculations demonstrated that the clustering of Zn cations is not energetically favored (as illustrated in the SI, **Figure S10**), confirming the homogeneity of Zn distribution and supporting the reported EDX measurements (**Figure 2**). In addition, the average exchange energy of all magnetic configurations was considered to determine the exchange interactions ($J_1$, $J_2$, and $J_3$) and Néel temperatures of the alloyed systems. Note that these results are valid for both bulk and monolayers, as interlayer interactions were shown to have minimal effect on the magnetic arrangement.[63]

The pristine $MnPS_3$ compound was used to calibrate the DFT model, including the computation of DFT+U energies for the four magnetic configurations outlined in **Figure 1(b)**, as detailed in the Methods below. The exchange interactions were then derived using **Eq. E1** (see Methods). For $MnPS_3$, the calculated exchange interactions were 0.64, 0.10, and 0.23 meV for $J_1$, $J_2$, and $J_3$, respectively, agreeing with previous studies.[22,40,64–66] Note that positive values of *J* indicate AFM interactions, while negative *J* designates FM interactions. The $T_N$ was determined using **Eq. E2** and found to be 114.3 K for the pristine $MnPS_3$. The observed deviation from the experimental value is attributed to a limitation of the mean-field approximation, often overestimating this temperature by at least 20%.[20] Exchange coefficients and Néel temperatures for each concentration are presented in **Table 1** and **Figure 6(c)**, respectively.

For x=0.75 compound, DFT results show that all the magnetic phases that exhibit AFM-Néel arrangement have the lowest energy, with mutual differences lower than 0.1 meV. Conversely, the energy difference between Néel and other magnetic configurations is substantially higher (>3 meV).

For Mn0.50, the analysis uncovered six magnetic phases with three FM and three AFM arrangements (**Figure M2**). Remarkably, all three AFM structures seem to have an AFM-stripy magnetic pattern, implying that the $Mn_xZn_{1-x}PS_3$ system undergoes a *dilution-dependent magnetic phase transition* around x=0.50. This phase transition aligns with the observed polarization in the PL of the Mn0.50 compound (**Figure S9**). Polarization can be expected for ferromagnetically coupled spin chains, where a strong coupling between the spin vector and the electric dipole oscillator was previously found.[27] The results indicate a preference for AFM-stripy configurations over the FM phases as the magnetic ground state, with an energy difference of more than 10 meV. Note that additional magnetic arrangements were found by doubling the rectangular unit cell of the x=0.50 compound (SI, **Figure S11**). Nevertheless, these arrangements are significantly higher in energy.

As presented in **Table 1**, there is a monotonic decline in the first and second exchange interactions with decreasing Mn content. The monotonic decrease in $J_1$ and $J_2$ is expected due to their strong dependency on the distance between Mn-Mn. However, an increase in $J_3$ is observed at x=0.50, likely associated with the Néel to stripy magnetic phase transition. Note that the observed weak magneto-optical coupling in the x=0.50 compound could be ascribed to the small value of $J_1$.

**Table 1.** Exchange parameters of the $Mn_xZ_{1-x}PS_3$ compounds obtained from DFT+U calculations.

| Compound \| x | $J_1$ | $J_2$ | $J_3$ |
|---|---|---|---|
| | | [meV] | |
| $MnPS_3$ \| 1.00 | 0.64 | 0.10 | 0.23 |
| $Mn_{0.75}Zn_{0.25}PS_3$ \| 0.75 | 0.54 | 0.05 | 0.15 |
| $Mn_{0.5}Zn_{0.5}PS_3$ \| 0.50 | 0.35 | 0.01 | 0.26 |

To further validate the averaged magnetic phase model, the $T_N$ values for the various compounds were calculated using equations provided in the Methods and plotted in **Figure 6(c)** (orange triangles). Note that the calculated Néel temperatures were normalized to the $T_N$ of $MnPS_3$ extracted from the SQUID measurements (79 K). As anticipated, the calculated $T_N$ demonstrates a gradual linear decrease with the increasing concentration of Zn(II), consistent with the $T_N$ extracted from the SQUID measurements. This confirms the reliability of the utilized average phases technique as a practical and effective approximation method to comprehensively characterize the magnetic behaviors of the $Mn_xZn_{1-x}PS_3$ alloyed compounds.

**Conclusions**

In conclusion, this work has demonstrated the tunability of the magneto-optical coupling in single crystals with the composition $Mn_xZn_{1-x}PS_3$. A vapor transport synthesis method provides the alloyed $Mn_xZn_{1-x}PS_3$, covering compositions from 0.37 ≤ x ≤ 1, encompassing the full spectrum of magnetic characters from paramagnetic attributes (x = 0.37) to the characteristic features of magnetic AFM-Néel materials (x = 1). A careful analysis of both absorption and photoluminescence (steady state- and transient-PL) of the alloyed crystals has revealed a distinct emission mechanism stemming from the $^4T_{1g} \rightarrow {}^6A_{1g}$ of the Mn(II) ion, which is partially allowed due to the hybridization of the S(p) and Mn(d) orbitals in the VB. The elucidation of the observed d-d transitions below the absorption edge relied on the assumption that the onset of the absorption edge emerging from deeper VB states in $MnPS_3$ based on DFT+U approach and previously published ARPES measurements, with the latter providing accurate positioning of the d-states within the PDOS. An origin of these peaks, visible in the pure and alloys systems, can be understood in terms of an oversimplified atomic picture of the 3d Mn states embedded in the p states coming from the S and P orbitals. Most notably, temperature-dependent PL intensity uncovered a correlation to the magnetic properties, showing a drop in the intensity near the crystal's Néel temperature. This correlation has been detected down to x=0.65 and was attributed to the persistence of spin-phonon coupling down to this concentration. Finally, the magnetic results were verified by DFT+U calculations based on an averaged phase model, allowing the extraction of the magnetic arrangements, spin-exchange parameters and the Néel temperatures of the alloyed compounds. Particularly, DFT calculations revealed a magnetic phase transition from AFM-Néel to AFM-stripy for the compound x=0.5. The present comprehensive work sheds light on the correlation between the optical and magnetic properties of bulk $Mn_xZn_{1-x}PS_3$ crystals, providing an understanding of the underlying physical phenomena leading to the connection between the magnetic and optical properties in the investigated AFM materials.

## Methods

**Materials.** Bulk single crystals of $Mn_xZn_{1-x}PS_3$ (0.25 ≤ x ≤ 1) were grown via the vapor sublimation synthesis. Briefly, stoichiometric amounts of Mn, red phosphorus, $S_8$, and Zn were ground in a mortar and pestle into a homogeneous mixture. The powder was transferred to a quartz tube and sealed under vacuum. Then, the tube was calcined in a gradient two-zone furnace for a week, where the hot (substrate zone) and cold (product zone) areas were set to 650˚C and 600˚C, respectively. After a week, the bulk $Mn_xZn_{1-x}PS_3$ single crystals were collected from the cold zone.

**PL measurements.** PL measurements were performed using a fiber-based confocal microscope embedded in a cryogenic system (attoDRY1000 closed cycle cryostat), with a heater capable of changing the temperature of the sample between 5 K – 300 K. The temperature was controlled by the Lake Shore PID system. The sample was excited by a 405/450 nm laser diode, and the emission was detected by the FERGIE spectrograph. On the optical head, a 473 nm long-pass dichroic mirror was used to transmit the laser to the sample while filtering it from the emission. An additional 800 nm long-pass filter was installed on the emission path to filter the laser completely. The lifetime measurements were acquired using a 450 nm pulse laser with a pulse duration of 70 ps. The transient PL was collected by a superconducting nanowire single photon (SNSPD) detector coupled to a PicoHarp300 time-correlated single photon counter (TCSP).

**SQUID.** SQUID measurements were performed in the Quantum Matter Research (QMR) center at the Technion Institute using a SQUID magnetometer Quantum Design MPMS3, which provides a sensitivity of ≤$10^{-7}$ emu. The magnetic susceptibilities were measured under an external magnetic field of 1000 Oe (0.1 Tesla), parallel or perpendicular to the lamellar planes of the crystals.

**Electron microscopy.** High-resolution scanning electron microscope (HR-SEM) images were acquired with Zeiss Ultra-Plus FEG-SEM. Energy Dispersive X-ray (EDX) spectra were gathered using Quanta 200 FEI E-SEM. Both measurements were conducted under an accelerating voltage of 20 kV. The EDX spectra were meticulously acquired to perform a quantitative assessment of the composition and to estimate the atomic ratios of [Mn]/[Zn] in the alloyed samples of $Mn_xZn_{1-x}PS_3$.

**Magnetic configurations of $Mn_xZn_{1-x}PS_3$.** The average magnetic phase analysis of the $Mn_xZn_{1-x}PS_3$ systems was performed to investigate the influence of Zn alloying on the template $MnPS_3$ compound. This approach involves using the magnetic configurations of the original $MnPS_3$ matrix with the rectangular unit cell (**Figure 1(b)**), substituting Mn with Zn based on the parameter x, and accounting for all non-equivalent phases. In this way, the unperturbed $MnPS_3$ compound serves as a reference in the calculations to calibrate the model. For each alloying concentration, the smallest supercell preserving the AFM arrangement was selected. This methodology transforms disordered systems into ordered models (supercells) by considering all possible non-equivalent magnetic arrangements within the employed supercell. Furthermore, for each alloying concentration, the lateral lattice parameters were optimized, keeping the position of the ions fixed.

Pure $MnPS_3$ (x=1.00). The exchange interactions in the isotropic compound can be derived from three linear equations, dependent on the DFT total energies of the meta-stable magnetic configurations, as shown in **Eq. E1**.

(Eq. E1)
$$J_1 = \frac{E_{FM} - E_{Néel} - E_{stripy} + E_{zigzag}}{8S^2}$$

$$J_2 = \frac{E_{FM} + E_{Néel} - E_{stripy} - E_{zigzag}}{16S^2}$$

$$J_3 = \frac{E_{FM} - E_{Néel} + 3E_{stripy} - 3E_{zigzag}}{24S^2}$$

Where $E_{FM}$, $E_{Néel}$, $E_{stripy}$, and $E_{zigzag}$ are the DFT total energies of the $MnPS_3$ magnetic configurations, and S is the spin magnetic moment.

Hence, the Néel temperature can be deduced from the magnetic configuration with the lowest energy, corresponding to the AFM- Néel arrangement, as demonstrated in **Eq. E2**:[67]

(Eq. E2)
$$T_N = \frac{S(S+1)}{3k_B}(-3J_1 + 6J_2 - 3J_3)$$

Where $k_B$ is the Boltzmann constant.

x=0.75. As previously mentioned, Zn atoms are randomly distributed within the honeycomb matrix. However, in an AFM environment, different Zn positions within the lattice generate different

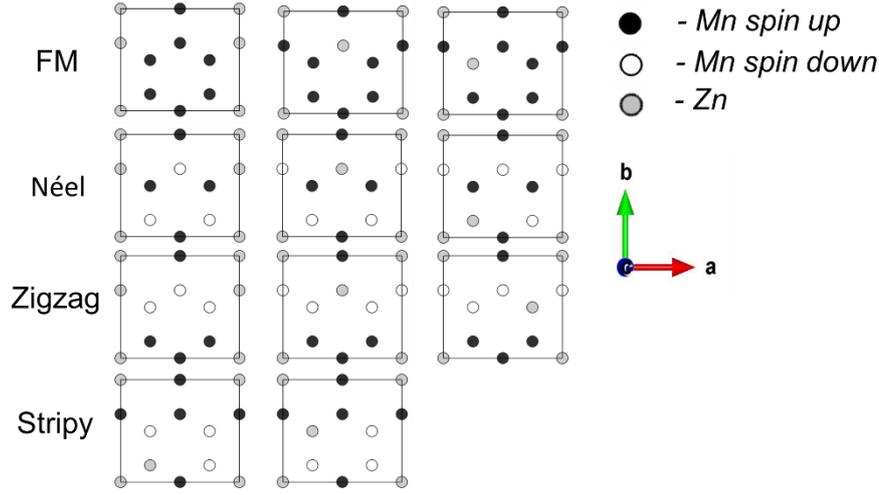

**Figure M1.** Magnetic phases of Mn$_{0.75}$Zn$_{0.25}$PS$_3$ (x=0.75).

exchange energies between the Mn atoms. Therefore, it is crucial to consider the average effect of all Zn displacements in the AFM crystal. To encompass all the magnetic phases with a 3:1 Mn to Zn ratio while maintaining zero net magnetization, a 2x1 rectangular unit cell template was used, where two atoms of Mn were replaced with Zn, as depicted in **Figure M1**. This yielded 11 non-degenerate virtual magnetic arrangements. Among these, the FM, AFM- Néel, and AFM-zigzag configurations contributed three magnetic arrangements each, while the AFM-stripy pattern contributed two. By averaging the exchange energy over a single magnetic cell, the <u>mean exchange energy</u> of a specific magnetic cell arrangement was obtained. The resulting exchange terms are then formulated as follows:

(Eq. E3)
$$J_1 = \frac{65E_{FM} - 69E_{Néel} - 70E_{stripy} + 74E_{zigzag}}{390S^2}$$

$$J_2 = \frac{325E_{FM} + 357E_{Néel} - 350E_{Stripy} - 332E_{zigzag}}{3900S^2}$$

$$J_3 = \frac{325E_{FM} - 318E_{Néel} + 1000E_{stripy} - 1007E_{zigzag}}{5850S^2}$$

Note that the coefficients of the DFT energies depend on the number of magnetic arrangements at each configuration (Néel, stripy, zigzag, FM). The magnetic ground state of the Mn0.75 compound was determined to be AFM-Néel. Consequently, the <u>average Néel exchange energy</u> was calculated according to the following equation:

(Eq. E4)
$$T_{\bar{N}} = \frac{S(S+1)}{3k_B}\left(-{}^{20}/_9 J_1 + 4J_2 - {}^{7}/_3 J_3\right)$$

Where the $\left(-{}^{20}/_9 J_1 + 4J_2 - {}^{7}/_3 J_3\right)$ term is the mean exchange energy of the Néel phases. An example of the calculation scheme for Mn$_{0.75}$Zn$_{0.25}$PS$_3$ is provided in the SI section 11.3.

**x=0.5.** Deploying the magnetic phases similarly to the x=0.75 case yields a degeneracy of the magnetic unit cell. Nevertheless, six distinct magnetic phases remained, as illustrated in **Figure M2**, comprising three FM and three AFM arrangements. In the three AFM arrangements, Mn ions are arranged in a stripy fashion and are therefore labeled as *Stripy1-3*.

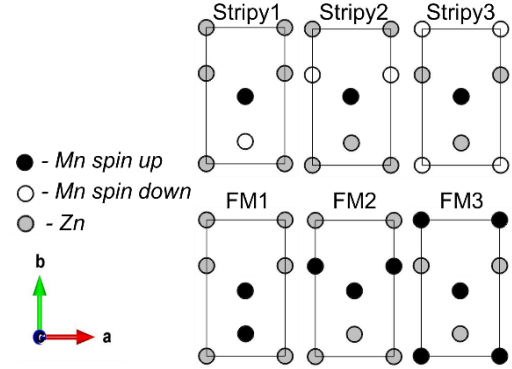

**Figure M2.** Non-degenerate magnetic phases of $Mn_{0.5}Zn_{0.5}PS_3$ (x=0.5).

To calculate the exchange interactions, the average energy of the FM phases was considered, given their almost identical energies, which results in the following equations:

(Eq. E5)
$$J_1 = \frac{1}{12} \cdot \frac{E_{stripy1} - 5 \cdot E_{stripy2} + E_{stripy3} + 3 \cdot E_{\overline{FM}}}{S^2}$$

$$J_1 = \frac{1}{12} \cdot \frac{E_{stripy1} - 5 \cdot E_{stripy2} + E_{stripy3} + 3 \cdot E_{\overline{FM}}}{S^2}$$

$$J_3 = \frac{1}{36} \cdot \frac{-11 \cdot E_{stripy1} + 7 \cdot E_{stripy2} + E_{stripy3} + 3 \cdot E_{\overline{FM}}}{S^2}$$

Where $E_{\overline{FM}}$ is the average DFT energy of the three FM arrangements. Importantly, the minor energy differences between the AFM-stripy phases result in coexisting magnetic phases at finite temperatures, making it challenging to determine which exchange energy should be considered to extract $T_N$. Therefore, the average exchange energy of the AFM-stripy magnetic phases was utilized, $\left(-J_1 + \frac{2}{3}J_2 - J_3\right)$, to extract $T_N$:

(Eq. E6)
$$\overline{T_N} = \frac{S(S+1)}{3k_B}\left(-J_1 + \frac{2}{3}J_2 - J_3\right)$$

**DFT.** For the calculations of the $Mn_xZn_{1-x}PS_3$ ground states, ab initio calculations were performed using the PBE (Perdew-Burke-Ernzerhof) generalized gradient exchange-correlation functional with PAW (projected augmented wave) pseudopotentials as implemented in the open-source Quantum Espresso package. All calculations were performed on a plane wave basis with a cutoff of 40Ry and 240Ry for the plane wave kinetic energy of the wavefunctions and charge density, respectively. We consider a mesh of 42 special k points in the irreducible wedge of the first Brillouin zone, corresponding to a grid of 10×8×1 in the Monkhorst Pack scheme. All calculations were spin-polarized. To account for the localization of the d orbital, we include the Hubbard potential (DFT +U), setting U to 3eV to achieve reasonable convergence of the Heisenberg parameters.[64] For structural optimization, we used ion relaxation posterior to a minimized Birch-Murnaghan plot, keeping the ratio of the a/b lattice parameters constant, as expected for the highly isotropic $MnPS_3$.[40,68] This approach should allow complete structural optimization of the $MnPS_3$ monolayers while preserving the symmetry of the crystal. All calculations refer to a monolayer slab with a vacuum length of 20Å. The density of states (DOS) have been calculated using VASP software.[69] The alloy has been considered within a supercell consisting of 4-fold primitive hexagonal unit cells. For each of the fractional concentrations (0.25, 0.50, 0.75), particular structural configurations have been considered. In order to improve the band gaps obtained within PBE exchange correlational functional, the modified Becke–Johnson (mBJ) potential[70] has been employed for $ZnPS_3$.[71] The convergence criteria for the energy and force were set to $10^{-7}$ eV and $10^{-3}$ eV/Å, respectively. A semi-empirical Grimme method[72] with a D3 parametrization (DFT-D3) was adopted[73] to account for dispersive forces.

**Acknowledgment**

E.L. thanks the support from the Deutsch – Israel Program (DIP, project no.NA1223/2-1), the Israel Science Foundation (ISF, project no. 2528/19) and the Binational Science Foundation – NSF (Project 2017/637). M.B. acknowledges support from the University of Warsaw within the project "Excellence Initiative-Research University" program. Access to computing facilities of PL-Grid Polish Infrastructure for Supporting Computational Science in the European Research Space and of the Interdisciplinary Center of Modeling (ICM), University of Warsaw are gratefully acknowledged. M.B acknowledges ACK Cyfronet AGH (Poland) for awarding this project access to the LUMI supercomputer, owned by the EuroHPC Joint Undertaking, hosted by CSC (Finland) and the LUMI consortium. T.W. acknowledges financial support from National Science Centre, Poland, under project no. 2023/48/C/ST3/00309. The authors thank Dr. Adam K. Budniak for his preliminary assistance in the preparation of the samples, and Dr. Kusha Sharma for her editing assistance.

**Supporting information**

Crystal's structure and composition determination, SQUID measurements, absorption spectra fit, PL under excitation at different laser wavelengths, PL lifetime analysis, temperature-dependent PL analysis, reproducibility of the measurements, Arrhenius model, electron-phonon coupling, polarized PL, first-principles calculations.